\documentclass{kapproc} 

\usepackage{procps} 

\usepackage[dvips]{graphicx}

\upperandlowercase

 \nochapequationnumber 

\normallatexbib 

\begin{document}

\articletitle{Nonlinear Quantum Dynamics}

\author{Salman Habib}
\affil{The University of California, Los Alamos National Laboratory,
Los Alamos, New Mexico, USA}  
\email{habib@lanl.gov}

\author{Tanmoy Bhattacharya,\altaffilmark{1} Andrew
  Doherty,\altaffilmark{2} Benjamin Greenbaum,\altaffilmark{3} Asa
  Hopkins,\altaffilmark{4} Kurt Jacobs,\altaffilmark{1,5} Hideo
  Mabuchi,\altaffilmark{4} Keith Schwab,\altaffilmark{6} Kosuke
  Shizume,\altaffilmark{7} Daniel Steck,\altaffilmark{1,8} and Bala
  Sundaram,\altaffilmark{9}}
   
\affil{\altaffilmark{1}Los Alamos National Laboratory,
  \altaffilmark{2}University of Queensland, \altaffilmark{3}Columbia
  University, \altaffilmark{4}California Institute of Technology,
  \altaffilmark{5}Griffith University, \altaffilmark{6}Laboratory for
  Physical Sciences, \altaffilmark{7}Tsukuba University,
  \altaffilmark{8}University of Oregon, \altaffilmark{9}City
  University of New York}

\begin{abstract}
The vast majority of the literature dealing with quantum dynamics is
concerned with linear evolution of the wave function or the density
matrix. A complete dynamical description requires a full understanding
of the evolution of measured quantum systems, necessary to explain
actual experimental results. The dynamics of such systems is
intrinsically nonlinear even at the level of distribution functions,
both classically as well as quantum mechanically. Aside from being
physically more complete, this treatment reveals the existence of
dynamical regimes, such as chaos, that have no counterpart in the
linear case. Here, we present a short introductory review of some of
these aspects, with a few illustrative results and examples.
\end{abstract}


\section{Introduction}

It is hard to imagine a scientific discipline older than the study of
dynamical systems. The remarkable history of the field testifies to
nature's inexhaustible store of subtlety and ability to surprise. Ever
since Galileo, remarkable experiments, deep theoretical insights, and
powerful calculational tools have all contributed to creating the rich
panorama that the field presents today.

From a theoretical perspective, dynamical systems are specified by the
rules of evolution and the physical objects to which these rules
apply. Our fundamental notions regarding both aspects have undergone
radical changes in the past few hundred years. Classical mechanics has
made way for quantum mechanics and absolute notions of space and time
have been replaced by the unified viewpoint of classical general
relativity. A key lesson to be drawn from these advances is that even
the most basic notions regarding the nature of physical information
must change as our overall understanding progresses.

The next step forward has yet to be taken: The clash between
relativity and quantum mechanics -- the choice between causality and
unitarity -- awaits resolution. However, on a less grand scale, the
tension between fundamentally different points of view is already
apparent in the discord between quantum and classical
mechanics. Unlike special relativity, where $v/c\rightarrow 0$
smoothly transitions between Einstein and Newton, the limit
$\hbar\rightarrow 0$ is singular. The symmetries underlying quantum
and classical dynamics -- unitarity and symplecticity, respectively --
are fundamentally incompatible with the opposing theory's notion of a
physical state: quantum-mechanically, a positive semidefinite density
matrix; classically, a positive phase-space distribution function.
 
Chaos provides an excellent illustration of this dichotomy of
world-views~\cite{peres}. Without question, chaos exists, can be
experimentally probed, and is well-described by classical
mechanics. But the classical picture does not simply translate to the
quantum view; attempts to find chaos in the Schrodinger equation for
the wave function, or, more generally, the quantum Liouville equation
for the density matrix, have all failed. This failure is due not only
to the linearity of the equations, but also the Hilbert space
structure of quantum mechanics which, via the uncertainty principle,
forbids the formation of fine-scale structure in phase space, and thus
precludes chaos in the sense of classical trajectories. Consequently,
some people have even wondered if quantum mechanics fundamentally
cannot describe the (macroscopic) real world.

It is therefore clear that there is more than sufficient motivation
for investigating the notion of nonlinearity in classical and quantum
theories. The main point of this article is to provide an angle of
vision which sets nonlinearity in its experimentally relevant
context. Familiar to control theorists~\cite{control} -- but much less
so to most physicists -- this perspective bridges the classical and
quantum points of view and smoothly connects them with each other.

The article is organized as follows. We will begin with a discussion
of the various possibilities of dynamical description, clarify what is
meant by ``nonlinear quantum dynamics,'' discuss its connection to
nonlinear classical dynamics, and then study two experimentally
relevant examples of quantum nonlinearity -- (i) the existence of
chaos in quantum dynamical systems far from the classical regime, and
(ii) real-time quantum feedback control.

The results described here are due to the efforts of many people
spread over the last thirty years or so, some results being even
older. Unfortunately, space limitations prevent anywhere near an
adequate job of referencing, for which a sympathetic understanding is
begged in advance. Restrictions also meant the omission of important
topics and explanations of derivations.

\section{Evolution: Isolated, Open, and Conditioned}

How should one describe a dynamical system? Before settling on a
definition, it is best to first ask some important physical
questions. As an illustrative example, a situation worth learning from
arose in the attempt to define a field-theoretic notion of a particle
in a general spacetime~\cite{wgu}. In Minkowski space, the formal
definition is simple: positive energy plane-waves, but this definition
cannot be extended to arbitrary metrics. It soon became clear that the
correct way to approach the problem was to give up the attempt to
arrive at a formal definition and replace it with a physical
definition: ``A particle is what a particle detector detects.'' Thus,
specifying a field theory Lagrangian is not sufficient to define the
notion of a particle, additionally we must model the detector and how
it couples to the field. Just what a physical ``particle'' is depends on
the design of the detector and the field-detector coupling.

Keeping the lesson of the above example in mind, we will explore three
different dynamical possibilities below: {\em isolated} evolution,
where the system evolves without any coupling to the external world,
{\em unconditioned open} evolution, where the system evolves coupled
to an external environment but where no information regarding the
system is extracted from the environment, and {\em conditioned open}
evolution where such information {\em is} extracted. In the third
case, the evolution of the physical state is driven by the system
evolution, the coupling to the external world, and by the fact that
observational information regarding the state has been obtained. This
last aspect -- system evolution {\em conditioned} on the measurement
results via Bayesian inference -- leads to an intrinsically nonlinear
evolution for the system state. The conditioned evolution provides, in
principle, the most realistic possible description of an
experiment. To the extent that quantum and classical mechanics are
eventually just methodological tools to explain and predict the
results of experiments, this is the proper context in which to compare
them.

\subsection{Isolated and Open Evolution}

Suppose we are given an arbitrary system Hamiltonian $H(x,p)$ in terms
of the dynamical variables $x$ and $p$; we will be more specific
regarding the precise meaning of $x$ and $p$ later. The Hamiltonian is
the generator of time evolution for the physical system state,
provided there is no coupling to an environment or measurement
device. In the classical case, we specify the initial state by a
positive phase space distribution function $f_{Cl}(x,p)$; in the
quantum case, by the (position-representation) positive semidefinite
density matrix $\rho(x_1,x_2)$ or, completely equivalently, by the
Wigner distribution function $f_W(x,p)$~\cite{wdf} (not positive).

The evolution of an {\em isolated} system is then given by the
classical and quantum Liouville equations for the {\em fine-grained}
distribution functions (i.e., the evolution is entropy-preserving):
\begin{eqnarray}
\partial_t f_{Cl}(x,p)&=& -\left[\frac{p}{m}\partial_x - \partial_x V(x)
\partial_p \right] f_{Cl}(x,p),\\
\label{cle}
\partial_t f_W(x,p)&=& -\left[\frac{p}{m}\partial_x
- \partial_x V(x) \partial_p \right] f_W(x,p) \nonumber\\
&& + \sum_{\lambda = 1}^\infty \frac{(\hbar/2i)^{2\lambda}}{(2\lambda
+ 1)!} \partial_x^{2\lambda+1}  V(x) \partial_p^{2\lambda+1} f_W(x,p),
\label{qle}
\end{eqnarray}
here we have assumed for simplicity that the potential $V(x)$ can be
Taylor-expanded. Note that these evolutions are both linear in the
respective distribution functions. Classically, the limit
$f_{Cl}(x,p)=\delta(x-\bar{x})\delta(p-\bar{p})$ is allowed, and, on
substitution in Eqn.~(\ref{cle}), yields Newton's equations. These may
then be interpreted as equations for the particle position and
momentum, although this identification is only formal at this stage
(as in the Minkowski space definition of a particle in the field
theory example). Quantum mechanically, this ultralocal limit is not
allowed as $f_W(x,p)$ must be square-integrable, therefore, even
formally, no direct particle interpretation exists (an obstacle that
arises as something new is added -- just like the difficulty with
generalizing the notion of a particle to arbitrary metrics discussed
above). 

The basic idea behind extension to open systems is simple to state but
not easy to implement in practice. The complete Hamiltonian now
includes a piece representing the environment and another, the
system-environment coupling. If the environment is in principle
unobservable, then a (nonlocal in time) linear master equation for the
system's reduced density matrix is derivable by tracing over the
environmental variables (in practice, tractable equations are
impossible to obtain without drastic simplifying assumptions such as
weak coupling, timescale separations, and simple forms for the
environmental and coupling Hamiltonians). In any case, the important
point to note is that the act of tracing over the environment does not
change the linear nature of the equations. Generally speaking, master
equations describing open evolution of {\em coarse-grained}
distributions augment the RHS of Eqns.~(\ref{cle}) and (\ref{qle})
with terms containing dissipation and diffusion kernels connected via
generalized fluctuation-dissipation relations~\cite{noneq}. While the
classical diffusion term vanishes in the limit of zero temperature for
the environment, this is not true quantum mechanically due to the
presence of zero-point fluctuations.

\subsection{Conditioned Evolution}

\begin{figure}
\begin{center}
   \includegraphics[width=8cm,height=6cm]{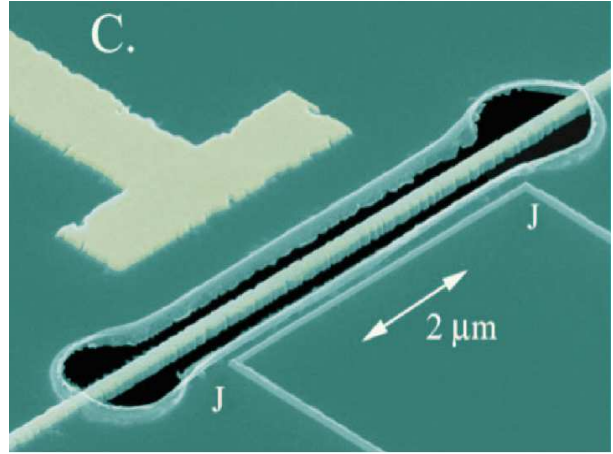}
   \caption[example]{ \label{set} A nanomechanical resonator: the thin
   central bar is coated with a conductor (gold) which also forms the
   T-shaped control electrode to the left. The thin line parallel to
   the resonator is the central island of a single-electron transistor
   which serves as the position sensor.}
\end{center}
\end{figure}

Conditioned evolution of the type we are interested in here is
fundamentally different from the equations discussed above. We assume
that measurements are possible on the environment and ask what the
evolution of the reduced density matrix of the system is, given that
the results of these measurements are known~\cite{contm}. Let us
consider an example. Suppose we wish to measure the position of a
nanomechanical oscillator (Fig.~\ref{set}). By electrostatically
coupling the resonator to a single-electron transistor (SET), and
measuring the (classical) SET current -- the measurement record -- we
are in fact measuring the transverse displacement of the resonator. In
this situation, the evolution of the reduced density matrix of the
system must contain a term that reflects the gain in information
arising from the measurement record (``innovation'' in the language of
control theory). This term, arising from applying a continuous analog
of Bayes' theorem, is intrinsically nonlinear in the distribution
function. The coupling to an external probe (and the associated
environment) will also cause effects very similar to the open
evolution considered earlier, and there can once again be dissipation
and diffusion terms in the evolution equations. The primary
differences between the classical and quantum treatments, aside from
the kinematic constraints on the distribution functions, are the
following: (i) the (nonlocal in $p$) quantum evolution term in
Eqn.~(\ref{qle}), and (ii) an irreducible diffusion contribution due
to quantum backaction reflecting the {\em active} nature of quantum
measurements.

We now consider a simple model of position measurement to provide a
measure of concreteness. In this model, we will assume that there are
no environmental channels aside from those associated with the
measurement. Suppose we have a single quantum degree of freedom,
position in this case, under a weak, ideal continuous
measurement~\cite{cmeqns}. Here ``ideal'' refers to no loss of
information during the measurement, i.e., a fine-grained evolution
with no change in entropy. Then, we have two coupled equations, one
for the measurement record $y(t)$,
\begin{equation}
dy = \langle x \rangle dt + dW/{\sqrt{8k}}
\label{record}
\end{equation}
where $dy$ is the infinitesimal change in the output of the
measurement device in time $dt$, the parameter $k$ characterizes the
rate at which the measurement extracts information about the
observable, i.e., the {\em strength} of the measurement~\cite{djj},
and $dW$ is the Wiener increment describing driving by Gaussian white
noise~\cite{noise}, the difference between the observed value and that
expected. The other equation -- the nonlinear stochastic master
equation (SME) -- specifies the resulting conditioned evolution of the
system density matrix, given below in the Wigner representation,
\begin{eqnarray}
f_W(x,p,t+dt)&=&\left[ 1 + dt\left[-\frac{p}{m} \partial_x  + \partial_x
V(x)\partial_p + D_{BA}\partial_p^2\right]\right. \nonumber\\
&& + \left.dt\sum_{\lambda = 1}^\infty \! \frac{\left(
\hbar/2i\right)^{2\lambda}}{(2\lambda + 1)!}  \partial_x^{2\lambda+1} 
V(x,t) \partial_p^{2\lambda+1} \right] f_W(x,p,t)
\nonumber \\
&& + dt\sqrt{8k}(x - \langle x \rangle)f_W(x,p,t)dW, 
\label{condq}
\end{eqnarray} 
where $D_{BA}=\hbar^2k$ is the diffusion coefficient arising from
quantum backaction and the last (nonlinear) term represents the
conditioning due to the measurement. In principle, there is also a
(generalized) damping term~\cite{mm}, but if the measurement coupling
is weak enough, it can be neglected. If we choose to average over all
the measurement results, which is the same as ignoring them, then the
conditioning term vanishes, but {\em not} the diffusion from the
measurement backaction. Thus the resulting linear evolution of the
coarse-grained quantum distribution is not the same as the linear
fine-grained evolution (\ref{qle}), but yields a conventional
open-system master equation. Moreover, for a given (coarse-grained)
master equation, different underlying fine-grained SME's may exist,
specifying different measurement possibilities.

The classical conditioned master equation [set $\hbar=0$ in
Eqn.~(\ref{condq}), holding $k$ fixed], 
\begin{eqnarray}
f_{Cl}(x,p,t+dt)&=&\left[ 1 - dt\left[\frac{p}{m} \partial_x  -
\partial_x V(x)\partial_p \right]\right]f_{Cl}(x,p,t) \nonumber\\
&& + dt \sqrt{8k}(x - \langle x \rangle)f_{Cl}(x,p,t)dW, 
\label{condc}
\end{eqnarray} 
does not have the backaction term as classical measurements are {\em
passive}: Averaging over all measurements simply gives back the
Liouville equation (\ref{cle}), and there is no difference between the
fine-grained and coarse-grained evolutions in this special case. [In
general, classical diffusion terms from ordinary open evolution can
also co-exist, as in the more general {\em a posteriori} evolution
specified by the Kushner-Stratonovich equation~\cite{kse}.] As a final
point, we will delay our discussion of how the classical trajectory
limit is incorporated in Eqn.~(\ref{condc}), i.e., the precise sense
in which the ``the position of a particle is what a position-detector
detects'' to the next section.

\section{QCT: The Quantum-Classical Transition}

As mentioned already, quantum and classical mechanics are
fundamentally incompatible in many ways, yet the macroscopic world is
well-described by classical dynamics. Physicists have struggled with
this quandary ever since the laying of the foundations of quantum
theory. It is fair to say that, even today, not everyone is satisfied
with the state of affairs -- including many seasoned practitioners of
quantum mechanics.

If quantum mechanics is really the fundamental theory of our world,
then an effectively classical description of macroscopic systems must
emerge from it -- the so-called quantum-classical transition (QCT). It
turns out that this issue is inextricably connected with the question
of the physical meaning of dynamical nonlinearity discussed in the
Introduction. The central thesis is that real experimental systems are
by definition not isolated, hence the QCT must be viewed in the
relevant physical context.

Quantum mechanics is intrinsically probabilistic, but classical theory
-- as shown above by the existence of the delta-function limit for the
classical distribution function -- is not. Since Newton's equations
provide an excellent description of observed classical systems,
including chaotic systems, it is crucial to establish how such a
localized description can arise quantum mechanically. We will call
this the {\em strong} form of the QCT. Of course, in many situations,
only a statistical description is possible even classically, and here
we will demand only the agreement of quantum and classical
distributions and the associated dynamical averages. This defines the
{\em weak} form of the QCT.

\subsection{The Strong Form of the QCT}

It is clear that the strong form of the QCT is impossible to obtain
from either the isolated or open evolution equations for the density
matrix or Wigner function. For a generic dynamical system, a localized
initial distribution tends to distribute itself over phase space and
either continue to evolve in complicated ways (isolated system) or
asymptote to an equilibrium state (open system) -- whether classically
or quantum mechanically. In the case of conditioned evolution,
however, the distribution can be localized due to the information
gained from the measurement. In order to quantify how this happens,
let us first apply a cumulant expansion to the (fine-grained)
conditioned classical evolution~(\ref{condc}), resulting in the
equations for the centroids ($\bar{x}\equiv\langle x\rangle$,
$\bar{p}\equiv\langle p\rangle$),
\begin{equation}
d\bar{x}=\frac{\bar{p}}{m}dt+\sqrt{8k}C_{xx}dW,
~~~d\bar{p}=\langle F(x)\rangle dt+\sqrt{8k}C_{xp}dW, 
\label{cumc}
\end{equation} 
where 
\begin{equation}
F(x)=-\partial_x V(x),~~~C_{AB}=(\langle AB\rangle + \langle
BA\rangle -2\langle A\rangle\langle B\rangle)/2, 
\end{equation} 
along with a hierarchy of coupled equations for the time-evolution of
the higher cumulants. These equations are the continuous measurement,
real-world, analog of the formal ultralocal Newtonian limit of the
distribution function in the classical Liouville
equation~(\ref{cle}). While Eqns.~(\ref{cumc}) always apply, our aim
is to determine the conditions under which the cumulant expansion
effectively truncates and brings their solution very close to that of
Newton's equations.  This will be true provided the noise terms are
small (in an average sense) and the force term is localized, i.e.,
$\langle F(x)\rangle=F(\bar{x})+\cdots$, the corrections being
small. The required analysis involves higher cumulants and has been
carried out in Ref.~\cite{bhj}. (Ref.~\cite{bhj} also points to
previous literature.) It turns out that the distribution will be
localized provided 
\begin{equation}
8k\gg\sqrt{\frac{(\partial_x^2 F)^2|\partial_x F|}{2mF^2}}
\end{equation}
and the motion of the centroid will effectively define a smooth
classical trajectory -- the low-noise condition -- as long as 
\begin{equation}
k\gg \frac{2|\partial_x F|}{S}
\end{equation}
where $S$ is the action scale of the system. Note that this condition
does not bound the measurement strength.

We now turn to the quantum version of these results. In this case, the
analogous cumulant expansion gives exactly the same equations for the
centroids as above, while the equations for the higher cumulants are
different. We can again investigate whether a trajectory limit
exists. Localization holds in the weakly nonlinear case if the
classical condition above is satisfied. In the case of strong
nonlinearity, the inequality becomes 
\begin{equation}
8k\gg\frac{(\partial_x^2 F)^2\hbar}{4mF^2}. 
\end{equation}
Because of the backaction, the low-noise condition is implemented in
the quantum case by a double-sided inequality:  
\begin{equation}
\frac{2|\partial_x F|}{s}\ll\hbar k\ll\frac{|\partial_x F|s}{4},
\label{newtonq}
\end{equation}
where the action is measured in units of $\hbar$, $s$ being
dimensionless. The left inequality is the same as the classical one
discussed above, however the right inequality is essentially quantum
mechanical. The measurement strength cannot be made arbitrarily large
as the backaction will result in too large a noise in the equations
for the centroids. As the action $s$ is made larger, both inequalities
are satisfied for an ever wider range of $k$. For continuously
measured quantum systems, trajectories that emerge in the macroscopic
limit follow Newton's equations, and hence can be chaotic as shown in
Ref.~\cite{bhj}. Thus, as speculated in a prescient paper by
Chirikov~\cite{chirikov}, measurement indeed provides the missing link
between ``quantum'' and ``chaos.'' 

\subsection{The Weak Form of the QCT}

If the conditions enforcing the strong form of the QCT are satisfied,
then the weak form follows automatically. The reverse is not true,
however: results from a coarse-grained analysis cannot be applied to
the fine-grained situation. Moreover, the violation of the strong
inequalities (\ref{newtonq}) need not prevent a weak QCT: It does not
matter if the distribution is too wide, as long as the classical and
quantum distributions agree, and, even if the backaction noise is
large, the coarse-grained distribution remains smooth and the weak
quantum-classical correspondence can still exist. Consequently, the
weak form of the QCT has to be approached in a different manner. In
fact, the weak version is just another way to state the conventional
decoherence idea; however, as discussed elsewhere~\cite{hjmrss}, mere
suppression of quantum interference does not guarantee the QCT even in
the weak form.

In a recent analysis carried out for a bounded open system with a
classically chaotic Hamiltonian, it has been argued that the weak form
of the QCT is achieved by two parallel processes~\cite{ghss},
explaining earlier numerical results~\cite{hsz}. First, the
semiclassical approximation for quantum dynamics, which breaks down
for classically chaotic systems due to overwhelming nonlocal
interference, is recovered as the environmental interaction filters
these effects. Second, the environmental noise restricts the foliation
of the unstable manifold (the set of points which approach a
hyperbolic point in reverse time) allowing the semiclassical
wavefunction to track this modified classical geometry.

It turns out that this analysis applies only to systems with a bounded
phase space. It is possible that topological restrictions on the
accessible phase space -- and not only the form of the particular
Hamiltonian -- play a crucial role in determining when the weak form
of the QCT actually applies. For example, this might explain why the
open-system quantum delta-kicked rotor is a counter-example to naive
expectations regarding the QCT~\cite{hjmrss}.

\section{Chaos and Quantum Mechanics}

The results of the previous section have already established that
classical chaos and quantum mechanics are not incompatible in the
macroscopic limit. The question then naturally arises whether observed
quantum mechanical systems can be chaotic far from the classical
limit? This question is particularly significant as closed quantum
mechanical systems are not chaotic, at least in the conventional sense
of dynamical systems theory~\cite{rice}. In the case of observed
systems it has recently been shown, by defining and computing a
maximal Lyapunov exponent applicable to quantum trajectories, that the
answer is in the affirmative~\cite{hjs}. Thus, realistic quantum
dynamical systems are chaotic in the conventional sense and there is
no fundamental conflict between quantum mechanics and the existence of
dynamical chaos.

\begin{figure}[t]
\begin{center}
   \includegraphics[width=8.5cm,height=9cm]{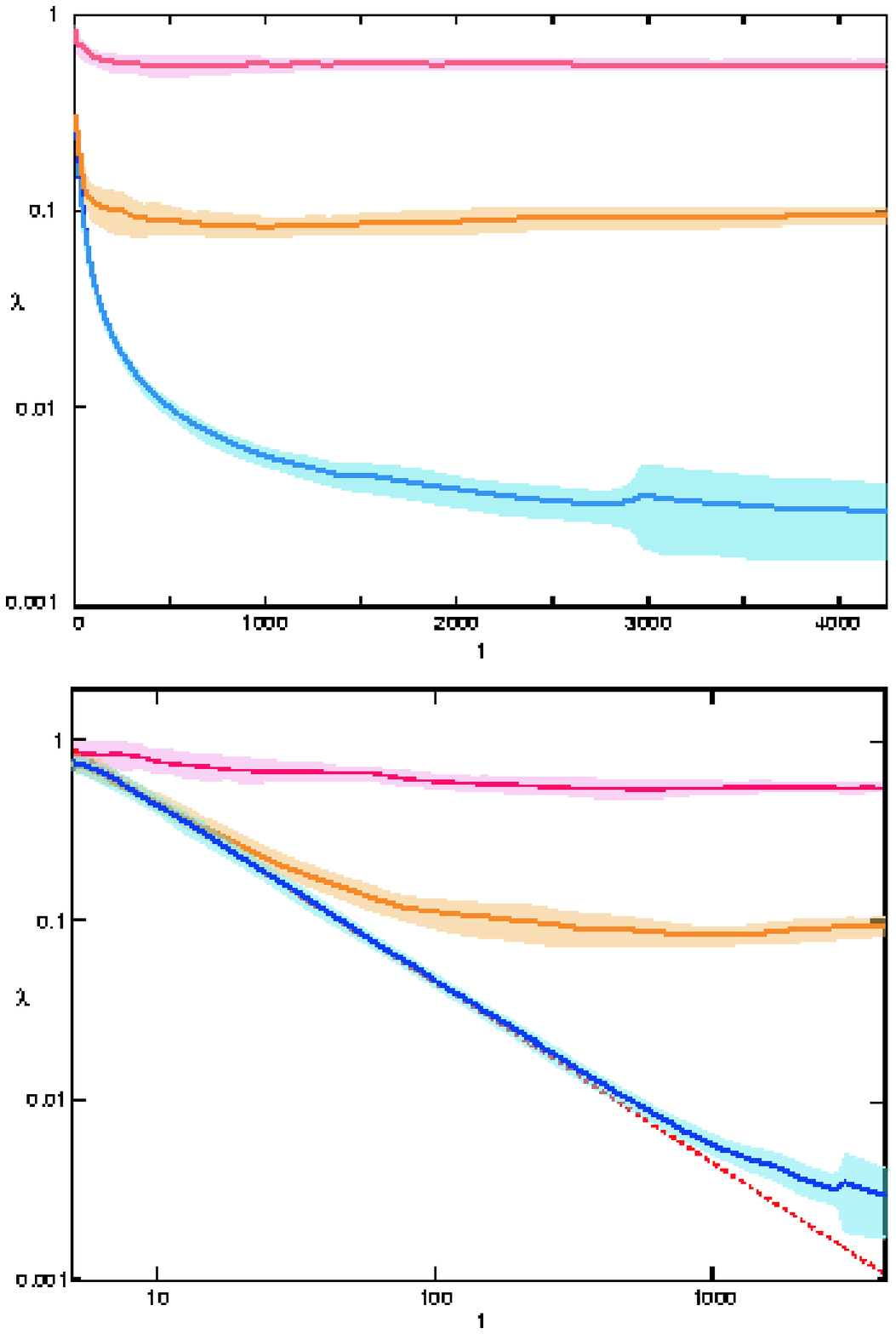}
   \caption[example]{ \label{lyapfig} Finite-time Lyapunov exponents
$\lambda(t)$ for a driven Duffing oscillator with measurement
strengths $k=5\times 10^{-4},~0.01,~10$, averaged over 32 trajectories
(linear scale in time, top, and logarithmic scale, bottom; bands
indicate the standard deviation over the 32
trajectories)~\cite{hjs}. The (analytic) $1/t$ fall-off at small $k$
values (dashed red line), prior to the asymptotic regime, is evident
in the bottom panel.}
\end{center}
\end{figure}

The basic idea in Ref.~\cite{hjs} is to focus attention on a single
time-series, say, the expectation value $\langle x\rangle$, and
analyze it for chaos. Following Ref.~\cite{hjs} the Lyapunov exponent
is defined to be
\begin{equation}
\lambda=\lim_{t\rightarrow \infty} \lim_{\Delta_s(0)\rightarrow
  0}\frac{1}{t}\ln\Delta_s(t)\equiv \lim_{t\rightarrow
  \infty}\lambda_s(t)
\label{lyap}
\end{equation}
where the subscript $s$ denotes the particular noise realization and
$\Delta(t)=|\langle x(t)\rangle - \langle x_{fid}(t)\rangle|$ defines
the divergence between two ``trajectories.'' The noise realization is
kept fixed when calculating $\Delta(t)$. For isolated systems it is
possible to prove that the Lyapunov exponent is zero, with the
finite-time exponent vanishing as $1/t$, as $t\rightarrow
\infty$~\cite{hjs}. This is consistent with our expectation of not
finding chaos for linear evolution. In the case of conditioned
nonlinear evolution, however, the situation can be dramatically
different as shown in Fig.~\ref{lyapfig}. What we find is that for
small $k$, $\lambda(t)$ first falls as $1/t$ (as for $k=0$), but then
stabilizes at an asymptotic value which is $k$-dependent, and
different from the classical value. Even at values of $k$ small enough
that the strong inequalities (\ref{newtonq}) are not satisfied,
$\lambda$ is finite, and the evolution is, thus, chaotic.

We stress that the chaos identified here is not merely a formal result
- even deep in the quantum regime, the Lyapunov exponent can be
obtained from measurements on a real system. Quantum predictions of
this type can be tested in the near future, e.g., in cavity QED and
nanomechanics experiments~\cite{exp}. Experimentally, one would use
the known measurement record to integrate the SME; this provides the
time evolution of the mean value of the position. From this fiducial
trajectory, given the knowledge of the system Hamiltonian, the
Lyapunov exponent can be obtained by following the procedure described
above. It is important to keep in mind that these results form only a
starting point for the further study of nonlinear quantum dynamics and
its theoretical and experimental ramifications.

\section{Quantum Feedback Control}

To illustrate an application of nonlinear quantum dynamics, we now
consider real-time control of quantum dynamical systems. Feedback
control is essential for the operation of complex engineered systems,
such as aircraft and industrial plants. As active manipulation and
engineering of quantum systems becomes routine, quantum feedback
control is expected to play a key role in applications such as
precision measurement and quantum information processing. The primary
difference between the quantum and classical situations, aside from
dynamical differences, is the active nature of quantum
measurements. As an example, in classical theory the more information
one extracts from a system, the better one is potentially able to
control it, but, due to backaction, this no longer holds true quantum
mechanically.

\begin{figure}
\begin{center}
   \includegraphics[width=8.5cm,height=6cm]{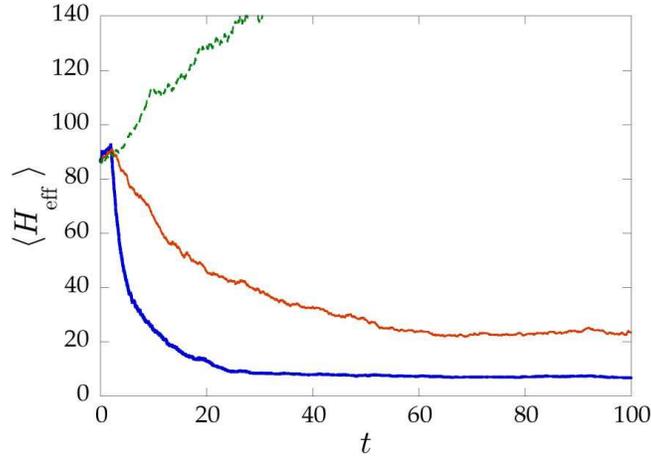}
   \caption[example]{ \label{atcool} Feedback cooling in cavity QED:
   Evolution of the mean atomic effective energy, with no cooling (top
   curve), cooling based on direct feedback of the photocurrent signal
   (middle), cooling based on feedback with a simple Gaussian state
   estimator (bottom). Note the improved cooling efficiency in the
   second case.}  
\end{center}
\end{figure}

Controlling quantum systems is possible using state-estimation ideas
as pioneered by Belavkin~\cite{dhjmt} or direct feedback of the
measured classical current~\cite{wm}. Applications studied so far
include controlling atomic~\cite{wmw} and qubit~\cite{rk} states as
well as active cooling of dynamical degrees of freedom. As one
example, let us consider an atom trapped in a high-finesse optical
cavity in the strong-coupling limit, with the output laser light
monitored via homodyne detection. The resulting photocurrent provides
information about the position of the atom in the cavity which, in
turn, can be used to cool the atom's position degree of freedom by
varying the intensity of the driving laser field~\cite{sjmbh}
(Fig.~\ref{atcool}). Nanomechanical resonators can also be cooled by
feedback. Here, the present state of the art has reached the point
where the resonators are less than a factor of 10 away from the
quantum limit, i.e., the point where the thermal energy is less than
the energy of the resonator ground state. Lowering the temperature of
the resonators to the mK regime would allow this goal to be
reached. In principle, active cooling could achieve this by measuring
the resonator position using a SET as described earlier
(Fig.~\ref{set}) and then applying (damping) feedback through the
control electrode~\cite{hjhs}. Experiments to test this idea are
currently in progress.

\begin{acknowledgments}
SH gratefully acknowledges the warm hospitality of the workshop
organizers, local scientists, and students at NDFI'04, as well as
the stimulating atmosphere of the meeting. This research is supported
by the Department of Energy, under contract W-7405-ENG-36. 
\end{acknowledgments}

\end{document}